\documentclass[12]{article}
\usepackage{amsmath,amssymb,amsthm}
\usepackage{mathtools}
\usepackage{mathrsfs}
\usepackage{bm}
\usepackage{slashed} 
\usepackage{multirow}
\usepackage{tikz}
\usepackage[utf8]{inputenc}
\usepackage{fullpage}
\usepackage{arydshln}
\usepackage{setspace}
\usepackage[margin=1in]{geometry}
\usepackage{dsfont}
\usepackage{graphicx}
\usepackage{cite}
\graphicspath{ {graphs/} }

\usepackage{hyperref}
\hypersetup{colorlinks=true,linkcolor=blue,anchorcolor=blue,citecolor=green,filecolor=blue
urlcolor=blue,bookmarksnumbered=true,pdfview=FitB}

\begin{document}
\title{\textbf{Effective operators in two-nucleon systems}}
\author{James P. Vary, Robert Basili, Weijie Du,  \strut\\ 
Matthew Lockner, Pieter Maris, Soham Pal and Shiplu Sarker}
\maketitle
\begin{center}
\textit{Department of Physics and Astronomy, Iowa State University, Ames, Iowa, U.S.A., 50011} 
\end{center}

\begin{abstract}
Effective Hamiltonians and effective electroweak operators are calculated with the Okubo-Lee-Suzuki 
formalism for two-nucleon systems.  Working within a harmonic oscillator basis, first without and then with
a confining harmonic oscillator trap, we demonstrate the effects of renormalization on observables calculated
for truncated basis spaces.  
We illustrate the renormalization effects for the root-mean-square point-proton radius, 
electric quadrupole moment, 
magnetic dipole moment, Gamow-Teller transition and neutrinoless double-beta decay operator using
nucleon-nucleon interactions from chiral Effective Field Theory.  Renormalization effects tend to
be larger in the weaker traps and smaller basis spaces suggesting applications to heavier 
nuclei with transitions dominated by weakly-bound nucleons would be subject to more significant 
renormalization effects within achievable basis spaces.
\end{abstract}

\section{Introduction}
Precision studies of electroweak properties of nuclei have become of great interest to complement 
major advances underway in experimental nuclear physics.  
As an example, significant experimental and theoretical efforts are aimed at searches 
for neutrinoless double-beta ($0 \nu 2 \beta$) decay which require significant investments in new experimental 
facilities and in theoretical advances. Our limited goal here is to use solvable two-nucleon systems 
within a configuration-interaction (CI) approach in order to explore the dependence 
of electroweak operators on the CI basis-space truncation when evaluating nuclear 
properties.  Information on the size of these effects can help interpret previous studies and
guide plans for calculations in larger nuclei.

We select systems of two nucleons interacting via realistic nucleon-nucleon ($NN$) interactions
both in free space and in a harmonic oscillator (HO) trap for investigating renormalization 
effects on a suite of electroweak properties.  
These systems are numerically solvable in a large HO basis space providing high precision 
results for comparison with approximate results.
This allows us to map out the effects arising from the correlations governed by different interactions, 
as well as the effects due to basis space truncation and the effects linked with the length scale 
of the environment, the trap.  To accurately calculate these effects 
we adopt the Okubo-Lee-Suzuki (OLS) method~\cite{Okubo:1954zz,Suzuki:1980yp,Suzuki:1982} to  
derive the basis truncation dependence of effective interactions and operators.  
By comparing matrix elements of these
derived effective operators with those from a truncated treatment of the original operators, 
we observe the truncation effects (conversely the renormalization effects) on each electroweak operator 
for each basis space and for each confining HO trap.
While most effects tend to be in the range of a few percent, the effects on the $0 \nu 2 \beta$
matrix element 
can be above a factor of two  
for some systems. Such cases suggest a careful treatment
of renormalization effects must be implemented for $0 \nu 2 \beta$ decay matrix elements 
in limited basis spaces.
Fortunately, these same OLS methods are adaptable to more realistic 
applications as in the case of the No-Core Shell Model 
(NCSM)~\cite{Stetcu:2004wh,Stetcu:2006zn,Lisetskiy:2009sh,Barrett:2013nh}
and to valence space effective interactions as well~\cite{Dikmen:2015tla}.

This work may be viewed in the context of a related pioneering work that extensively 
investigated the deuteron and its electromagnetic form factors with renormalization 
group methods \cite{Anderson:2010aq}.
The importance of maintaining consistency in the renormalization of both the Hamiltonian and
all other observables is the theme we share. The renormalization method, interactions, observables and  presence of a trap distinguish our work from Ref. \cite{Anderson:2010aq}.

\section{Theoretical Framework}
\subsection{Many-Body Systems}
We seek to solve a Hamiltonian $H$ eigenvalue problem expressed in a suitable
basis and, once the eigenvectors are obtained, to evaluate matrix 
elements of additional observables $O$.  For nuclear physics applications, such as the NCSM,
the resulting matrix for $H$ is infinite dimensional.  With truncation, the matrix of $H$ becomes
numerically tractable, allowing the study of results as a function of the
finite basis parameters in order to estimate the converged results and their uncertainties.  
For the NCSM, we express $H$ in terms of the relative kinetic energy
operator $T_{\rm rel}$ acting between all pairs of nucleons in the $A$--nucleon system
and an interaction term $V$ that may include multi-nucleon interactions as
\begin{eqnarray}
H &=& T_{\rm rel} +V. \
\end{eqnarray}

By adopting a complete basis of Slater determinants, $|\Phi_j \rangle$ for $A$ nucleons, 
developed from a chosen single-particle basis, we express the complete space problem as a matrix 
eigenvalue problem.   That is, the eigenvalues $E_k$ and eigenstates $|\Psi_k \rangle$, 
expanded in our complete basis of Slater determinants, obey the equations 
\begin{eqnarray}
H |\Psi_k \rangle\ &=& E_k |\Psi_k \rangle \\
|\Psi_k \rangle &=& \sum_{j} A_{kj}  |\Phi_j \rangle, 
\end{eqnarray}
where $A_{kj}$ denotes the expansion coefficient.

The selection of the complete single-particle basis is flexible but we will follow a popular
choice and adopt the three-dimensional HO due to its well-studied analytical properties 
that facilitate numerical applications and the retention of the underlying symmetries 
of $H$~\cite{Barrett:2013nh}. We take the neutron
and proton mass $m$ to be the same ($938.92$ MeV, their average measured mass), 
so that the only length scale in the HO single-particle basis can be expressed in terms of the HO energy 
$\hbar\Omega$ as
\begin{eqnarray}
b &=& \sqrt{\frac{\hbar}{m\Omega}}. 
\label{Eq:HOlengthscale}
\end{eqnarray} 

\subsection{Finite Matrix Truncation Approach}
In practical applications, it is advantageous to define the many-body truncation with  
$N_{\rm max}$, the maximum of the total HO quanta in the retained
Slater determinants above the minimum total HO quanta for the $A$ nucleons \cite{Barrett:2013nh}.  
A quantum of the HO single-particle state is twice the radial quantum number $n$ 
plus the orbital quantum number $l$. That minimum total HO quanta for the $A$ nucleons 
also depends on the number of neutrons $N$ and protons $Z$ that comprise the system.
Zero is the minimum total HO quanta for the two-nucleon systems addressed in this work.

We define the $P$--space (or ``model space'') as the basis space retained by this $N_{\rm max}$ 
truncation.  The infinite-dimensional space beyond this $N_{\rm max}$ truncation 
is called the $Q$--space.
For a sufficiently large $N_{\rm max}$, some observables are seen to converge in
very light nuclei for interactions which do not couple strongly to high momentum states and 
when computational resources are sufficient.  For example,
using a chiral N$^2$LO $NN$ interaction~\cite{Epelbaum:2014sza, Epelbaum:2014efa}, 
the ground state (gs) energy of $^6$Li has been calculated~\cite{Maris:2016wrd,Binder:2015mbz,Binder:2018pgl} 
in a sequence of HO basis spaces.  With extrapolation to the complete basis,
the result is $-31.0(2)$ MeV~\cite{Binder:2018pgl}.
 Here, the parenthesis specifies the uncertainty as $200$ keV in the extrapolation.  
The basis space truncation for the largest finite basis 
employed in the extrapolation is $N_{\rm max}=18$.  
At  $N_{\rm max}=18$ and $\hbar\Omega = 28$ MeV the gs energy, 
which is also a variational upper bound of the exact result, 
is already $-29.928$ MeV, i.e., about $1.1$ MeV above the extrapolated result. 
For comparison, the experimental gs energy is $-31.995$ MeV \cite{Tilley:2002vg}.  

However, other observables, such as the root-mean-square (rms) point-proton radius
and electric quadrupole transitions, converge poorly up through 
$N_{\rm max}=18$~\cite{Shin:2016poa}.
For long-range observables such as the rms point-proton 
radius, the theoretical results are insufficiently converged
to provide directly a meaningful comparison with experiment.  
For example, with extrapolations, the rms point-proton radius for $^6$Li, 
has significant uncertainties~\cite{Shin:2016poa}.

For all these reasons, it may be advantageous to soften the interactions and to promote
improved convergence of the eigenvalue problem.  As we explain 
in the next section, this softening, or renormalization of the interaction, 
also necessitates renormalizing the operators corresponding
to these other observables.  That is, we need to consistently 
derive the effective operators for all observables in the chosen model space.

\subsection{Effective Hamiltonian and Operators}
Once the complete basis space and the $P$--space are defined, we can address the
development of an effective Hamiltonian $H_{\rm eff}$ for the $P$--space that
formally retains a subset of the eigenvalues of the complete space.  We adopt the 
OLS method~\cite{Okubo:1954zz,Suzuki:1980yp,Suzuki:1982} which we briefly
outline here. More details, including specifics for including three-nucleon interactions,
are found in Ref.~\cite{Barrett:2013nh}.

The formal structure 
of the OLS approach is visualized by first considering $H$ 
in the complete basis space
and defining the unitary transformation $U$ that diagonalizes $H$ to produce the Hamiltonian's spectral form
$H_d$ along with the $P$--space projection of $U$, which we denote $W^P$.
\begin{eqnarray}
H_d \mathrel{\mathop:}= UHU^\dagger\\
W^P \mathrel{\mathop:}= PUP.
\end{eqnarray}  

Provided that the projected eigenvectors form a complete set of linearly independent vectors in the $P$--space, 
we can then construct a finite basis transformation $\widetilde{U}^P$, which is unitary in the $P$--space, as
\begin{eqnarray}
\widetilde{U}^P \mathrel{\mathop:}= \frac{W^P}{\sqrt{W^{P\dagger} W^P}}.
\end{eqnarray} 

With this transformation, we define our $H_{\rm eff}$ for the $P$--space with
\begin{eqnarray}
H_{\rm eff}  \mathrel{\mathop:}= \widetilde{U}^{P\dagger} H_d \widetilde{U}^P \\
     =  \widetilde{U}^{P\dagger} U H U^\dagger \widetilde{U}^P.
\end{eqnarray} 

We denote the OLS transformation $U_{\rm OLS}$ by the combination
\begin{eqnarray}
U_{\rm OLS} \mathrel{\mathop:}= \widetilde{U}^{P\dagger} U.
\end{eqnarray} 

We define our $P$--space to accommodate the lowest set of eigenvalues of
the original Hamiltonian though other choices are feasible, such as retaining states whose
eigenvectors have the largest probabilities of $P$--space configurations.
We furthermore note that $H_{\rm eff}$ is not unique since there is the freedom of a residual
$P$--space unitary transformation that preserves $H_d$.  
Additional mathematical
issues have been addressed in Ref.~\cite{Viazminsky:2001_JPV} 
such as the breakdown when linearly-dependent projected eigenvectors are encountered.
We did not encounter this breakdown in the calculations reported in this work.  
A central issue for the
current work is to investigate the effects of the corresponding transformation on
the observables $O$ needed to generate consistent renormalizations.  That is,
we define consistent effective operators
\begin{eqnarray}
O_{\rm eff} \mathrel{\mathop:}= U_{\rm OLS} O U_{\rm OLS}^\dagger
\end{eqnarray}
for calculations of observables with the $P$--space eigenfunctions of $H_{\rm eff}$.
\footnote{Note that we are addressing a $P$--space which generally includes
eigenstates with different conserved quantum numbers. The $U_{\rm OLS}$
transformation will properly manage scalar operators that conserve the
symmetries of $H$ as well as non-scalar operators that may induce transitions
between eigenstates.
}

While these steps provide the formal framework, the essential question of a practical 
implementation requires further discussion.  In the NCSM, one introduces an auxiliary
confining potential, which is later removed, and solves for the OLS transformations on
a subset of the nucleons in the nucleus (typically two or three nucleons) in what is dubbed
a ``cluster approximation''~\cite{Navratil:2000ww,Barrett:2013nh}.  The derived few-nucleon
effective interaction is designed to renormalize the strong few-nucleon correlations in the
presence of other nucleons approximated by the auxiliary potential.
This effective interaction is subsequently employed to define an $A$-nucleon effective Hamiltonian. 
The cluster approximation is guaranteed to produce the exact results as either the cluster 
size is increased to reach the full $A$-nucleon system or as the $P$--space truncation
is removed.  
The two-nucleon cluster approximation of the NCSM serves as a paradigm for 
introducing and solving the $NN$ systems of this work.  Our aim is to investigate the
ramifications  
of this approach for effective electroweak operators derived in a manner 
consistent with the softened interaction.

To this stage, we have described the formal structure of the OLS method and we have discussed
its applications within the NCSM.  For the two-nucleon systems we address below, either without or
with a HO trap, we apply the OLS approach in the relative coordinate system.
Since the $NN$ Hamiltonian is defined with the conserved symmetries of each $NN$ partial wave
(channel), we apply the OLS method to each $NN$ channel independently.  
That is, we solve for individual OLS transformations
in a relative HO basis of fixed total angular momentum, coupled spin, parity and charge 
-- the conserved quantum numbers for each $NN$ channel.  Then, we calculate the
effective non-scalar operators with OLS transformations from the $NN$ channels
required for that operator.
In these applications,
the length scale of Eq. (\ref{Eq:HOlengthscale}) is defined with the reduced mass
which we take to be $469.46$ MeV for all $NN$ channels.

\section{Applications to Two-Nucleon Systems}
\subsection{Deuteron Ground State}
We define an initial system to consist of two nucleons described by the $H$
of Eq. (1) whose gs energy we investigate in this subsection.  Motivated by the NCSM cluster approximation
framework, we will define a second two-nucleon system in the next subsection
that adds a confining HO interaction (trap). 
Both systems are numerically solvable. We refer to the 
numerical solutions of these two systems as their ``exact'' results and
we present these results in the Appendix.  Using graphical representations, we compare 
these exact results with solutions from a truncation approach and with solutions from an
OLS effective Hamiltonian approach.  We refer to the results from the truncation approach 
and the results from the OLS approach each as ``model'' results.

We adopt $NN$ interactions from chiral Effective Field Theory (EFT) and we include the Coulomb interaction
between proton pairs in the Hamiltonian.
Specifically, we employ the $NN$ interactions of the Low Energy Nuclear Physics International Collaboration (LENPIC) \cite{Epelbaum:2014sza, Epelbaum:2014efa,Maris:2016wrd,Binder:2015mbz,Binder:2018pgl} which have been developed for each chiral order up through N$^4$LO. These LENPIC interactions employ a semilocal 
coordinate-space regulator and we select the interactions with the regulator range 
$R=1.0$ fm \cite{Binder:2015mbz,Binder:2018pgl}. 
We refer to these interactions as ``LENPIC--X'' where ``X'' defines the specific chiral order (LO, NLO, N$^2$LO, N$^3$LO or N$^4$LO). We also employ the chiral EFT interaction of Ref.~\cite{Entem:2003ft} with momentum-space regulator $500$ MeV which we refer to as ``Idaho--N$^3$LO''.  All of these chiral EFT interactions are charge-dependent.

We consider the neutron-proton ($np$) system since the deuteron gs provides the only bound $NN$ state.  
In particular, we solve
for the deuteron gs energy for each interaction using three approaches.
First, we obtain a high-precision result by diagonalizing the Hamiltonian in the 
coupled $^3S_1-^3D_1$ channel in a very large HO basis ($N_{\rm max}=400$)
for three different values of the HO energy $\hbar\Omega$.  We have 
verified that these gs energies produce the same result as the numerical solution of the Schroedinger
equation to at least 5 significant figures \cite{Epelbaum:2014sza, Epelbaum:2014efa} in all cases. 
We refer to these results from diagonalization at $N_{\rm max}=400$ as the ``exact'' results.  One may
view these exact results as creating a discretized approximation to the continuum
and we note that the largest eigenvalues exceed 1 GeV in all cases investigated here. 
Second, we solve for the gs energy in the HO bases truncated at lower $N_{\rm max}$ values 
to produce model results for the simple truncation approach.  Third, we solve for the OLS effective
Hamiltonian at each value of $N_{\rm max}$ and $\hbar\Omega$, following the methods 
described above, to produce the model results for the OLS approach.   
We follow this same approach for the neutron-neutron ($nn$) and proton-proton ($pp$) 
channels needed for some of the transitions addressed in this work.

\begin{figure}[ht!]
\centering
\includegraphics[width=15.65cm]{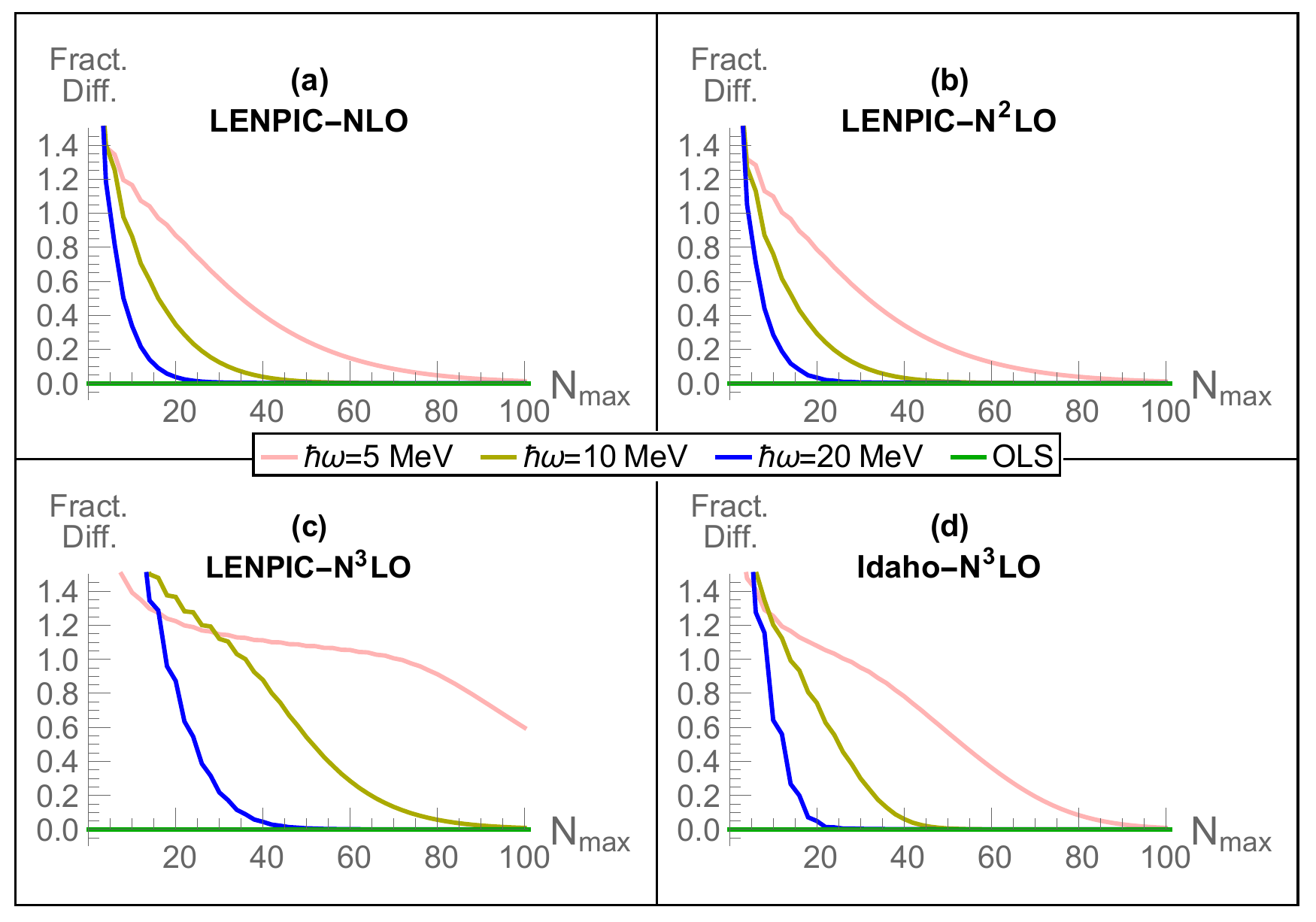}
\caption{The fractional differences, where Fract. Diff. of an observable 
$=(model-exact)/|exact|$,
for the deuteron gs energy at three values of the HO energy $\hbar\Omega$ as a function of the $P$--space 
limit $N_{\rm max}$.  The model results from diagonalizing the $P$--space truncated Hamiltonian matrix produce Fract. Diff. curves that decrease towards zero with increasing  $N_{\rm max}$ in accordance with the variational principle.  The model results from diagonalizing the OLS--renormalized Hamiltonian matrix reproduce the exact results at each $N_{\rm max}$ to high precision, yielding flat and overlapping green lines for their Fract. Diff. plots in all cases.  
Panels (a), (b) and (c) correspond to the Hamiltonians constructed with the 
LENPIC chiral EFT interactions \cite{Epelbaum:2014sza, Epelbaum:2014efa,Maris:2016wrd,Binder:2015mbz,Binder:2018pgl}  
at NLO, N$^2$LO and N$^3$LO, 
respectively.  We employ the LENPIC interactions with coordinate-space regulator $R=1.0$ fm. 
Panel (d) 
corresponds to the Hamiltonian constructed with the Idaho--N$^3$LO potential \cite{Entem:2003ft} with momentum-space regulator $500$ MeV. 
}
\label{fig:Heff}
\end{figure}

The model results of the truncation and OLS approaches are used to calculate their
fractional difference with respect to our exact results for each observable 
(i.e., observables calculated with the $N_{\rm max}=400$ wave functions) where the
fractional difference, Fract. Diff., is defined as the scaled difference $(model-exact)/|exact|$. 
The Fract. Diff. results for the deuteron gs energy are presented as curves in Fig.~\ref{fig:Heff}
for a representative selection of our $NN$ interactions.   
We do not show the results for LENPIC-LO since they are similar to the
LENPIC--NLO results. Also, we do not show the LENPIC--N$^4$LO results which are similar
to those of LENPIC--N$^3$LO.

From the results in Fig.~\ref{fig:Heff}, we observe that the convergence rates for the truncation
approach can depend significantly on the chiral order with the LENPIC $NN$ interaction.  In particular,
there is a dramatic slowing of the convergence rates at N$^3$LO  
as seen in 
panel (c) compared to panels (a) and (b). 
The dependence of the convergence rates on 
the LENPIC chiral orders is also revealed in many-body
observables where slower convergence leads to larger extrapolation 
uncertainties \cite{Maris:2016wrd,Binder:2015mbz,Binder:2018pgl}.

Comparing the LENPIC $NN$ interaction results in Fig.~\ref{fig:Heff} also reveals 
changing shapes of the convergence patterns with increasing chiral order 
for the bases using HO energy $\hbar\Omega = 5~{\rm and}~10$ MeV.  In particular, 
the case with $\hbar\Omega = 5$ MeV develops a region
showing significantly reduced slope, nearly a plateau, with increasing $N_{\rm max}$ at N$^3$LO  
(panel (c)).
The results for the Idaho--N$^3$LO interaction shown in 
panel (d)
indicate convergence patterns intermediate 
to LENPIC--N$^2$LO (panel (b)) and LENPIC--N$^3$LO (panel (c)).
These regions of reduced slope correspond to
Fract. Diff $\approx 1.0$ which defines a region of $P$--spaces where the lowest 
solution is transitioning between an unbound state and
a bound state with increasing $N_{\rm max}$.  Thus, for some interactions 
at lower values of $\hbar\Omega$, we observe a plateau-like behavior as seen in panels
(c) and (d)
of Fig.~\ref{fig:Heff}.  
In these same cases, after crossing over to a bound state solution, the convergence rate accelerates.
We have investigated this plateau and have found a correlation with a changing feature of the wavefunction:
while the solution is moving across the plateau with increasing $N_{\rm max}$ it is building up its $d$--state 
component from near zero to near its final value.  When it nearly acquires its final value, the energy decreases 
to a bound state (Fract. Diff. falls below 1.0) and accelerates its convergence rate.  We examined 
other interactions exhibiting this plateau in Fract. Diff. and found a similar correlation 
with buildup of the $d$--state probability.
We anticipate that, at sufficiently low values of the basis $\hbar\Omega$, we would find a similar 
plateau with all realistic interactions for the deuteron.

The dependence of the convergence rate on the basis $\hbar\Omega$ for the truncation approach in 
Fig.~\ref{fig:Heff} is systematic -- from most rapid convergence at $\hbar\Omega = 20$ MeV to slowest at $\hbar\Omega = 5$ MeV.  
We have not sought to optimize the choice of $\hbar\Omega$ though
that is often a point of interest in many-body applications.  Our interest here is rather to feature results for a range of
choices of $\hbar\Omega$ that will be useful for our investigation of electroweak observables in the system
of the following subsection.

The Fract. Diff. for the results of the OLS approach in Fig.~\ref{fig:Heff} always remains at zero, to within our numerical
precision, which is what one anticipates.  That is, since the OLS approach should provide the exact gs energy in any 
basis space, these results serve as a verification of our numerical procedures.  The OLS procedure
reproduces that subset of eigenvalues of the complete problem compatible with the dimensionality fixed
by $N_{\rm max}$, including eigenvalues lying high in the continuum. We verify the accuracy of the
eigenvalues from the OLS approach by direct comparison with the corresponding subset of results 
from the complete problem for each set of $P$--space basis parameters.  
We have confirmed that our OLS eigenvalues agree with the respective subset of the exact eigenvalues to 
at least 6 significant figures.

\begin{figure}[ht!]
\centering
\includegraphics[width=15.65cm]{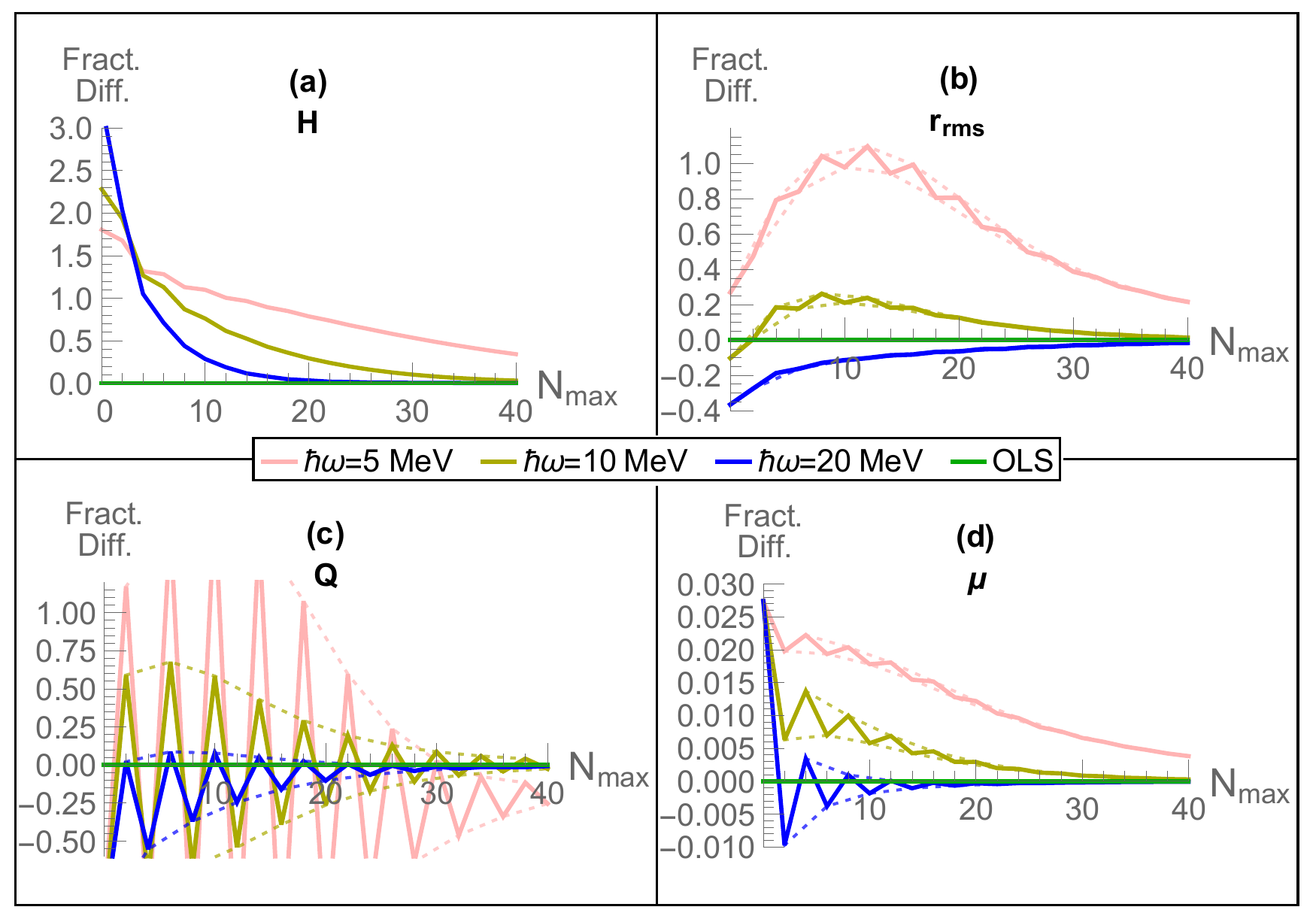}
\caption{The fractional differences 
for a selection of deuteron properties at three values of the HO energy basis parameter $\hbar\Omega$ as a function of the $P$--space 
limit $N_{\rm max}$.  Following the scheme of Fig. \ref{fig:Heff}, we present the Fract. Diff. for the truncated basis calculations 
(three colored curves approaching zero at high $N_{\rm max}$) and for the OLS renormalized calculations 
(green curves all coincident with zero).
All results are obtained with the LENPIC--N$^2$LO interaction with regulator $R=1.0$ fm. 
The gs energy in panel (a) is an expanded version of the gs energy in panel (b) of Fig. \ref{fig:Heff}.
The rms point-proton radius $r_{\rm rms}$ appears in panel (b), the electric quadrupole moment 
$Q$ in panel (c) 
and the magnetic moment $\mu$ in panel (d).  
The model results using the OLS transformation method reproduce the exact results at each $N_{\rm max}$ 
to high precision, yielding flat and overlapping green lines for their Fract. Diff. plots in all cases.  
The short dashed lines provide envelopes for results with sawtooth patterns.}
\label{fig:Deuteron_obervables}
\end{figure}

We now present additional observables for the deuteron that represent baseline results 
for later comparison.
In Fig. \ref{fig:Deuteron_obervables} we show the gs energy (panel (a)), rms point-proton radius
(panel (b)), electric quadrupole moment (panel (c)) and magnetic moment (panel (d)) for the same set of 3 
HO basis parameters as in Fig. \ref{fig:Heff} using the LENPIC--N$^2$LO interaction.  
Here, we expand the scale to see details of the results 
from $N_{\rm max}$ = 0 -- 40. Clearly, the results in the truncated basis exhibit strong deviations from
their exact values and those deviations exhibit non-smooth behavior, such as sawtooth patterns, 
with increasing $N_{\rm max}$.  
The excursions in the electric quadrupole moment in panel (c) are especially prominent.  These features
represent the role of an $s$--state plus $d$--state combination that are added with each increase 
in $N_{\rm max}$ by 2 units.  With the addition of two such combinations (increase $N_{\rm max}$ by 4 units)
we include states with canceling asymptotic tails.  This observation helps one to understand 
why results differing by 4 units in $N_{\rm max}$ follow a simpler trend in Fig. \ref{fig:Deuteron_obervables} 
-- a trend visualized by the dotted line connecting the successive maxima and another dotted line connecting
the successive minima of the sawtooth patterns.  Such a visualization is also applied to the maxima and the
minima of the sawtooth patterns of the other observables in Fig. \ref{fig:Deuteron_obervables}, 
the rms radius (b) and magnetic moment (c).  Note that the scale for the magnetic moment is greatly  enlarged relative to the other scales in Fig. \ref{fig:Deuteron_obervables} indicating it is rather insensitive 
to basis truncation effects.

Here again, the calculations of the effective operators with the OLS method,
when employed with the OLS wave functions in the same $P$--space, provide the exact results 
to within 6 digits for each observable at each value of the trap.  The OLS results are seen 
as the flat green lines at Fract. Diff. $\approx 0$ in Fig. \ref{fig:Deuteron_obervables}.  

\subsection{Two Nucleons in HO Trap -- Ground State Energy}
For the purpose of investigating effective electroweak operators and gauging a range of representative 
behaviors anticipating future applications to finite nuclei, we augment the initial system of the
previous subsection with the addition of a confining interaction or trap.  We adopt a HO confining potential 
and separate 
the effects of the trap into center-of-mass and relative motion.  Since the center-of-mass motion factorizes,
its wavefunction is an exact HO eigenstate and not affected by the $NN$ interaction acting
in relative coordinates. We subsequently consider only the relative motion of the two nucleons 
in the confining HO potential of relative motion and employ the Hamiltonian:

\begin{eqnarray}
H &=& T_{\rm rel} +V + \frac{1}{2} \mu\Omega^2 r^2  \
\end{eqnarray}
For these systems, we adopt HO energies for the confining interaction which match the basis values of 
$\hbar\Omega = 5,~10,~{\rm and}~20$ MeV introduced above.

In this second system, the HO trap simulates the role of the nuclear medium in which the
two-nucleon subsystem is embedded.  For applications
where we expect weakly bound nucleons to dominate the electroweak properties, we provide
results with the confining strength parameter $\hbar\Omega = 5$ MeV.  
Then we progress to simulate contributions from moderately bound
to deeply bound nucleon pairs with $\hbar\Omega = 10~{\rm and}~20$ MeV, respectively.  
We also note that the addition of the HO trap emulates the mathematical framework for the 
two-body cluster approximation in the NCSM where the HO trap is added
during the development of the effective Hamiltonian and then removed at the
stage of defining the effective interaction~\cite{Barrett:2013nh}.  Such a treatment of
the HO trap as a pseudo-potential is known to improve convergence in NCSM applications
with the OLS approach using the cluster approximation.
We therefore anticipate that studying the two-nucleon system with various HO traps, including its electroweak
properties, will provide insights into renormalization effects on observables in future NCSM applications
to finite nuclei.

\begin{figure}[ht]
\centering
\includegraphics[width=16cm]{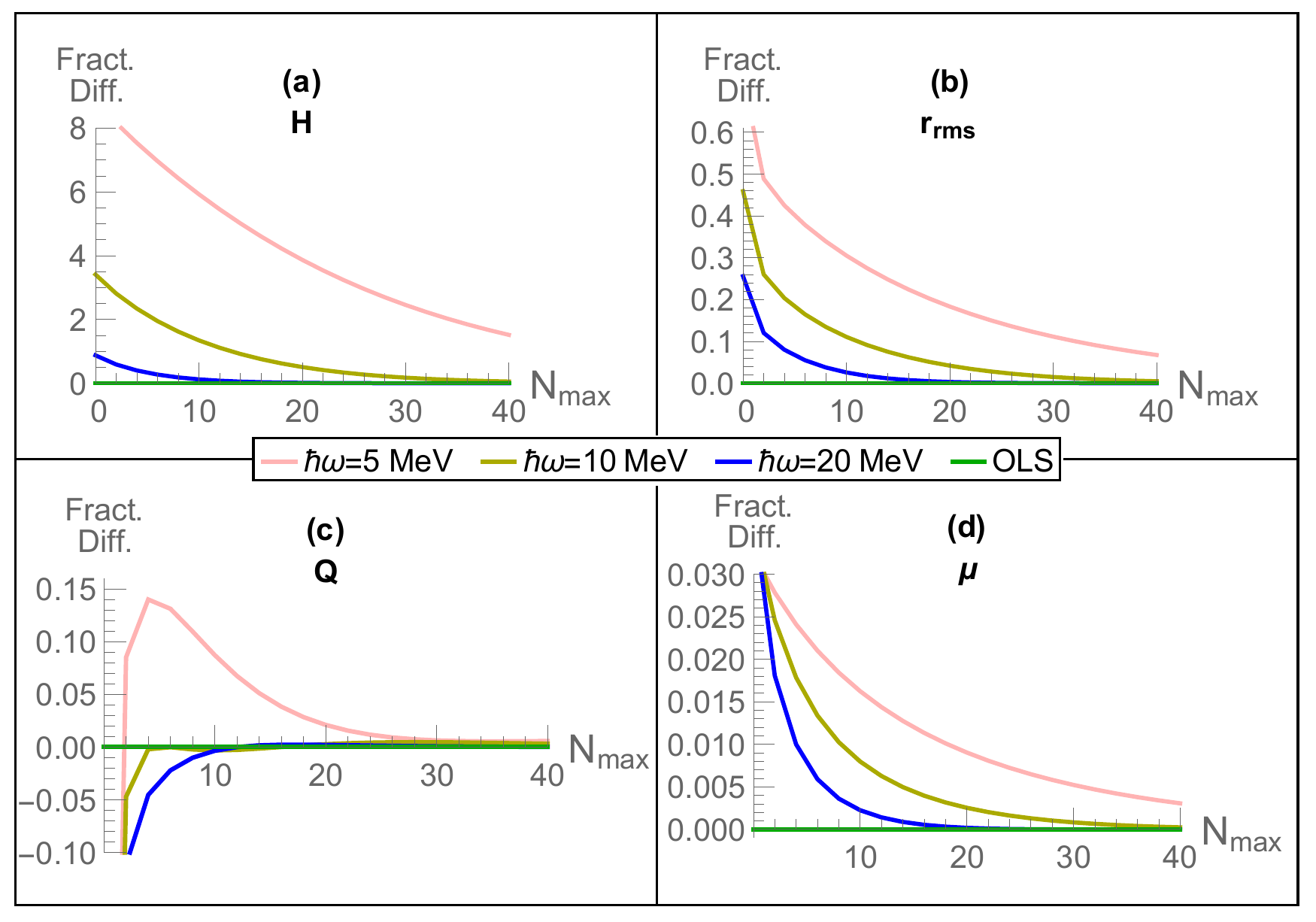}
\caption{Fractional differences between model and exact results as a function of the $P$--space for selected gs observables of the two-nucleon system in the $^3S_1 - ^3D_1$ channel for three different HO traps. 
The $NN$ interaction is the LENPIC--N$^2$LO $NN$ interaction with regulator $R=1.0$ fm. HO energies $\hbar\Omega$ for the bases 
correspond to the HO energies of the traps. The observables correspond to the eigenenergy (a), the rms point-proton radius (b), the electric quadrupole moment (c) and the magnetic dipole moment (d). 
}
\label{fig:staticMoments}
\end{figure}

For simplicity, we elect to retain only a subset of the LENPIC $NN$ interactions for this second system.  
In panel (a) of Fig. \ref{fig:staticMoments} we show the Fract. Diff. for the gs energy in the $^3S_1-^3D_1$ 
channel for the three different traps.  Again, we list the exact results in the Appendix. Note that, for this 
second system, the exact results depend on the HO strength parameter $\hbar\Omega$.
The convergence rate for results of the truncation approach is again systematic -- slowest with $\hbar\Omega = 5$ MeV
and fastest with $\hbar\Omega = 20$ MeV.  However, the scale for Fract. Diff. in panel (a) 
of Fig. \ref{fig:staticMoments} is much larger than in Fig. \ref{fig:Heff}. Nevertheless, the results of the OLS approach
again provide agreement with the exact gs energy results over all choices of $P$--space as evident by
the coincident flat green lines at Fract. Diff. = 0.

\subsection{Two Nucleons in HO Trap -- Electromagnetic Observables}
We now turn our attention to additional observables in this second system where we again aim
to compare the results of the truncation approach with those of the OLS approach.  We note that a consistent
OLS treatment of various long-range observables has been investigated in the past and 
results of truncation versus OLS approaches have been shown to be in the range from
a few percent to about ten percent~\cite{Stetcu:2004wh, Stetcu:2006zn, Lisetskiy:2009sh}.  Our aim
here is to investigate both electromagnetic and weak observables over a wide range
of basis spaces but in the limited two-nucleon system.  In the end, we find large effects 
for matrix elements of some observables in cases with traps having small HO energies and/or 
in cases having small $N_{\rm max}$ values.  
This suggests that truncation effects are more severe and renormalization effects are more important 
when approaching observables involving weakly bound states and transitions involving resonance 
states.

For the initial set of gs observables beyond the gs energy, we again examine the rms point-proton
radius $r_{\rm rms}$, quadrupole moment $Q$ and
magnetic dipole moment $\mu$.  We note that the role of the harmonic confining interaction is significant 
since the wave functions will therefore have gaussian asymptotic properties which
moderate the long-range contributions of these operators.  In true nuclear bound state environments, 
the asymptotic wave functions are exponential and provide significant contributions to the 
long-range observables.

For both the truncation approach and the OLS approach, we again
present the results for the gs expectation value of these operators as a fractional difference 
from the exact results in Fig. \ref{fig:staticMoments}.  The exact results are, again, those obtained in 
$N_{\rm max} = 400$ calculations and are given in the Appendix for completeness.
Note the major differences in the scales of Fig. \ref{fig:staticMoments}.
The results of truncation are the largest for the gs energy (see discussion above) and then decreasing 
in size for $r_{\rm rms}$, $Q$ and $\mu$ in that order.  The results of truncation are also largest 
at $\hbar\Omega = 5$ MeV 
and smallest at $\hbar\Omega = 20$ MeV.
Physically, one expects that, as $\hbar\Omega$ increases, the trap will become more important than the 
$NN$ interaction in the Hamiltonian.  
With this decreasing role of the $NN$ interaction, results from the truncation approach will trend towards 
the simpler situation of two nucleons in a simple HO Hamiltonian where they are increasingly
representative of the exact results with the same simple Hamiltonian. 
Indeed, the main impression of the results of the truncation approach with a trap in 
Fig.  \ref{fig:staticMoments} is the generally smooth
trends that contrast the results without a trap presented in Fig. \ref{fig:Deuteron_obervables}
-- in particular, the sawtoooth behavior seen in Fig.  \ref{fig:Deuteron_obervables} 
is absent from Fig. \ref{fig:staticMoments}.
While $r_{\rm rms}$ (panel (b)) and $\mu$ (panel (d)) appear to have a monotonic convergence pattern, 
just like the gs energy, the convergence pattern for $Q$ is slightly more complicated.  
The Fract. Diff. for $Q$ in the truncation 
approach at lower HO trap energies exhibits a sign change
while it exhibits a tendency towards monotonic behavior at larger trap energies.

As shown previously in Fig. \ref{fig:Deuteron_obervables}, we find that, as expected, 
the results for the gs observables 
in Fig. \ref{fig:staticMoments} within the OLS approach agree with the exact gs results over all choices of
$P$--spaces and traps.  This agreement is evident through the Fract. Diff. remaining
zero (flat green lines) for the OLS results in all panels.  
We checked that the Fract. Diff. for the OLS results shown in Fig. \ref{fig:staticMoments} are zero 
to at least six significant figures. 

\subsection{Two Nucleons in HO Trap -- Weak Observables}
\begin{figure}[ht]
\centering
\includegraphics[width=16cm]{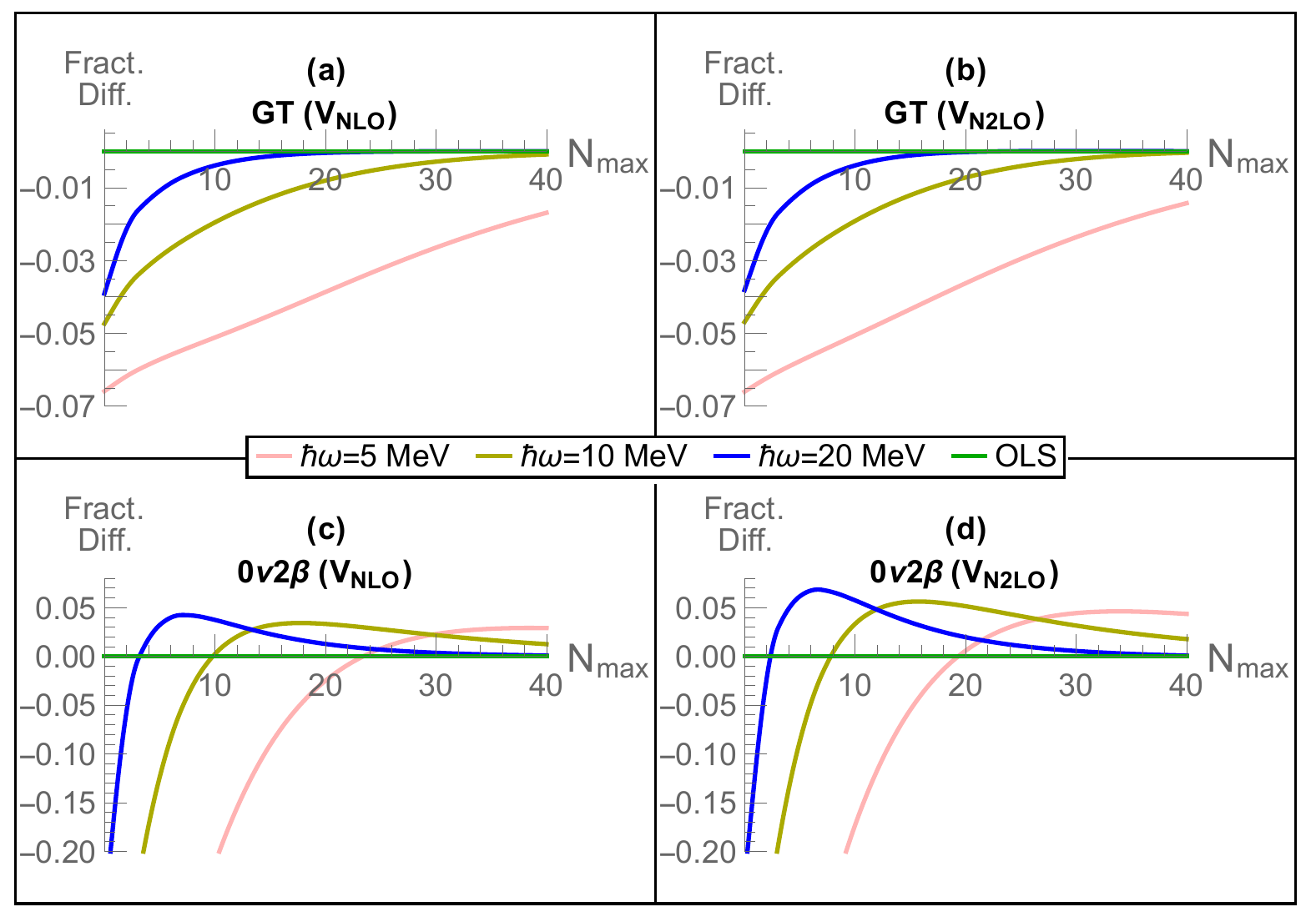}
\caption{Fractional differences between model and exact results as a function of the $P$--space for selected gs transitions from the 
lowest state of the $^1S_0$ $nn$ system in three different HO traps.  Panels (a) and (b) are for the allowed  
GT-transition to the gs 
of the $^3S_1 - ^3D_1$ channel. Panels (c) and (d) are for the $0\nu 2\beta$-decay to the gs of the $^1S_0$  $pp$ system. The $NN$ interaction for cases (a) and (c) are taken to be the LENPIC--NLO potential, while we adopt the LENPIC--N$^2$LO potential 
for cases (b) and (d). All results shown employ LENPIC $NN$ interactions with coordinate space 
regulator $R = 1.0$ fm.  
}
\label{fig:transitions}
\end{figure}

We now consider the Gamow-Teller (GT) beta decay matrix element for the transition from the gs of the $^1S_0$ $nn$ 
channel to the gs of the $^3S_1 - ^3D_1$ deuteron channel shown in panels (a) and (b) of Fig. \ref{fig:transitions} for the same traps as above. 
For panel (a) (and also for panel (c)) we
adopt the LENPIC--NLO $NN$ interaction. For panel (b) (and also for panel (d)) we
adopt the LENPIC--N$^2$LO $NN$ interaction. The GT operator is the simple spin-isospin form~\cite{BohrMottlesonVol1}. 
Our aim is to explore the role of two different, but lower, chiral order interactions with panels (a) and (b) (and also panels (c) and (d)). Future works will employ GT operators from chiral EFT
where the emphasis will shift to having all observables, including the eigenvalues, obtained with chiral EFT operators at the same
chiral order.

\begin{figure}[ht]
\centering
\includegraphics[width=16cm]{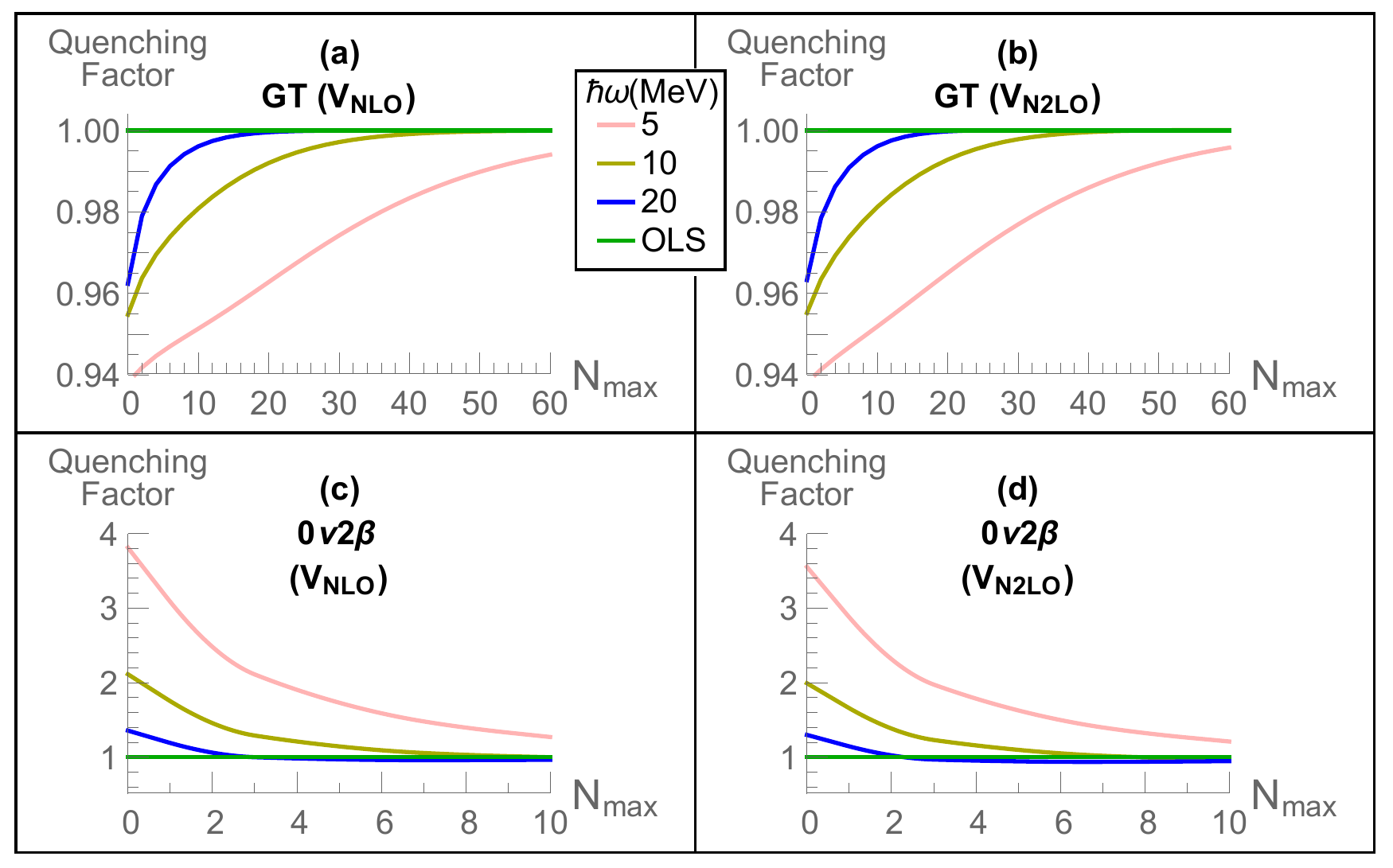}
\caption{Quenching factor defined as Exact/Model for GT-decay and $0\nu 2\beta$-decay 
matrix elements as a function of the $P$--space for gs transitions from the 
lowest state of the $^1S_0$ $nn$ system in three different HO traps.  Panels (a) and (b) are for the allowed  
GT-transition to the gs of the $^3S_1 - ^3D_1$ channel. Panels (c) and (d) are for the $0\nu 2\beta$-decay to the gs of the $^1S_0$  $pp$ system. The $NN$ interaction for cases (a) and (c) are taken to be the LENPIC--NLO potential, 
while we adopt the LENPIC--N$^2$LO potential for cases (b) and (d). 
All results shown employ LENPIC $NN$ interactions with coordinate space regulator $R = 1.0$ fm. 
A quenching factor greater than unity signals an enhancement of the model results is required 
to arrive at the exact results. 
}
\label{fig:quenching}
\end{figure}

The Fract. Diff. for the GT matrix elements in panels (a) and (b) of of Fig. \ref{fig:transitions} 
displays convergence patterns 
in the truncation approach similar to the operators in panels (a), (b) and (d) of Fig. \ref{fig:staticMoments} 
but on a reduced scale and with the opposite sign.  
The exact result, whose overall sign is arbitrary and has no physical consequence, 
is negative with our numerical procedures (see Table 4 in the Appendix). 
With increasing $P$--space dimension, the truncation results  
rise towards the exact results from below producing the Fract. Diff. trends for the GT matrix elements.
Thus, in smaller basis spaces, the results of the truncation approach for this gs-to-gs transition would require the application 
of a scaling, or quenching factor, that is less than unity in order to match the exact results.  
In order to help make this point more clear, we present plots of the quenching factor (defined as the Exact/Model result) 
in Fig. \ref{fig:quenching} for the results in Fig. \ref{fig:transitions}.  We note in panels (a) and (b) of Fig. \ref{fig:quenching}
that the quenching factor in truncated calculations of the GT matrix element deviates 
from unity by, at most, about 6\%.  The largest deviations from unity and the slowest convergence rates are found with the weakest 
HO trap energy of  $\hbar\Omega = 5$ MeV where the $nn$ and $np$ ground states are spread in coordinate space compared
with their distributions in the other HO traps. This systematic decrease  
of the GT quenching factor from unity with increasing spatial distribution
of the two-nucleon system is reminiscent of the approximate phenomenological decrease in the quenching factor for GT matrix elements in valence spaces with increasing atomic number~\cite{MartinezPinedo:1996vz}.
On the other hand, as expected, the OLS approach again produces the exact GT results 
for all our choices of the HO trap energies and for all choices of $P$--space 
(flat green lines at zero (unity) for panels (a) and (b) of Fig. \ref{fig:transitions} (Fig. \ref{fig:quenching})).

Finally, owing to intense current interest, we investigate the $0\nu 2\beta$-decay operator 
within the same approaches and exhibit the results for three traps in panels (c) (with LENPIC--NLO) 
and panel (d) (with LENPIC--N$^2$LO) of Fig. \ref{fig:transitions} and Fig. \ref{fig:quenching}.   
We adopt the $0\nu 2\beta$-decay 
operator from chiral EFT of Ref. \cite{Prezeau:2003xn}, which is consistent with chiral N$^2$LO.
For calculating our exact results, we employ a basis space with $N_{\rm max}=200$ which is sufficient 
for high-accuracy calculations with the LENPIC interaction at N$^2$LO.
We elect not to apply the LENPIC semilocal coordinate space regulator to this $0\nu 2\beta$-decay
operator as this regulator is not gauge invariant. 
  
The $0\nu 2\beta$-decay results in panels (c) and (d) of Fig. \ref{fig:transitions} and Fig. \ref{fig:quenching} 
reveal that convergence patterns of the truncation approach are sensitive to the HO energy of the trap.  
Furthermore, in contrast with several results seen above, 
the $0\nu 2\beta$-decay Fract. Diff. results of Fig. \ref{fig:transitions} do not 
appear to approach the exact result  
with increasing $P$--space dimension until we reach 
about $N_{\rm max}=(10,20,40)$ for $\hbar\Omega = (20,10,5)$ MeV respectively. 
At $N_{\rm max}=0$, the magnitude of the results from the truncation approach 
are much smaller than the exact results; 
with increasing $N_{\rm max}$ the results from the truncation approach increase but overshoot the 
exact results by about 5\% before converging towards the exact results.
That is, at small $N_{\rm max}$ values, the $0\nu 2\beta$-decay results with the truncation approach are significantly suppressed indicating large effects due to evaluating matrix elements of this operator in truncated basis spaces.
For more details of this lower region of $N_{\rm max}$, we turn to panels (c) and (d) of Fig. \ref{fig:quenching}.

The quenching factors in the truncated calculations of $0\nu 2\beta$-decay in panels (c) and (d) 
of Fig. \ref{fig:quenching} rise significantly above unity in the low $N_{\rm max}$ region contrary 
to what we find for the GT matrix elements.  The rise is largest with the weakest trap strength of 
$\hbar\Omega = 5$ MeV and smallest with the strongest trap of $\hbar\Omega = 20$ MeV.
We interpret these results to indicate that significant contributions from intermediate range components 
of the $0\nu 2\beta$-decay operator are omitted with truncations to the smaller model spaces.
Furthermore, these results suggest that truncated calculations of $0\nu 2\beta$ transition 
matrix elements between weakly bound nucleons are likely to require 
quenching factors greater than unity (i.e., enhancements) in small model spaces. 
In other words, these substantial excursions 
above unity may have significant implications for eventual applications in finite nuclei.  
It is also worth remarking that the OLS transformation is different for
the initial and final states of both the GT and $0\nu 2\beta$ matrix elements.  Therefore, obtaining the exact 
results for these matrix elements for all $P$--spaces by the OLS method is another non-trivial test 
of our numerical procedures. 

\section{Summary and Outlook}
Our initial application to the deuteron ground state revealed the order-of-magnitude effects of
simple $P$--space basis truncations compared with exact results as a function of the basis 
HO energy $\hbar\Omega$ for a set of realistic $NN$ interactions.  The smallest 
value studied, $\hbar\Omega=5$ MeV, produced the largest truncation effects (hence the largest 
renormalization effects) for these interactions.  We also showed that, for a wide range 
of $P$--spaces and a selection of $NN$ interactions, the Okubo-Lee-Suzuki approach 
consistently reproduced the exact results which one anticipates from a theoretical perspective.

Effective Hamiltonians and effective electroweak operators were then calculated for
two nucleons in a confining harmonic oscillator trap as a function of the $P$--space.
For this system, matrix elements of all OLS-derived effective operators again agree
with exact results in all model spaces and for all traps investigated.
We quantify the deviations of the simple truncated space results 
from the exact results for three different traps as a function of the $P$--space.
We illustrate these effects for the root-mean-square point-proton radius, 
electric quadrupole moment, 
magnetic dipole moment, Gamow-Teller transition and neutrinoless double-beta decay operators using
$NN$ interactions from chiral Effective Field Theory.  

From the results shown in Figs. \ref{fig:staticMoments} - \ref{fig:quenching}, we found that  
the size of error in the matrix element of an observable introduced by a truncated 
basis space approach depends on the observable, the value of the HO parameter of the trap 
$\hbar\Omega$, and the severity of the $P$--space truncation given by $N_{\rm max}$. 
Long-range observables, such as the 
root-mean-square point-proton radius and the electric quadrupole moment 
exhibited larger errors due to truncation to smaller spaces than the magnetic dipole 
and GT operators. 
We also found surprisingly large truncation errors for the double beta decay operator -- larger than that of $Q$, and of the same order as the $r_{\rm rms}$ radius.  On the other hand,  the GT operator exhibited behaviors
similar to the magnetic moment as may be expected.

While these results appear to be reasonable in the
qualitative sense and are consistent with previous investigations, the quantitative
dependences shown here may be useful in estimating uncertainties for observables obtained 
in truncated many-body calculations with realistic $NN$ interactions.  In particular, 
since renormalization effects tend to be larger in cases with weaker traps and smaller basis spaces, 
applications to heavier nuclei, for both transitions between weakly bound nucleons and 
to continuum states, will likely be subject to the more significant renormalization effects.  

The results presented here also signal the 
approximate magnitude of the corrections that the OLS renormalization provides for 
each of our selected observables.  These corrections, which can be obtained with 
OLS renormalization, should be carried forward to the appropriate many-body applications. 
It will also be important to implement the chiral Effective Field Theory treatment of the electroweak 
operators that are consistent with the chiral Effective Field Theory of the strong inter-nucleon interactions.

\section{Acknowledgements}
We acknowledge fruitful discussions with Hugh Potter, Evgeny Epelbaum, Hermann Krebs 
and Jacek Golak.
This work was supported in part by the US Department of
Energy (DOE) under Grant Nos.~DE-FG02-87ER40371, DE-SC0018223 (SciDAC-4/NUCLEI) 
and DE-SC0015376 (DOE Topical Collaboration in Nuclear Theory for Double-Beta Decay 
and Fundamental Symmetries). 
PM thanks the Funda\c{c}\~{a}o de Amparo\`{a} Pesquisa do Estado de S\~{a}o Paulo (FAPESP) for support under grant No 2017/19371-0.
Computational resources were
provided by the National Energy Research Scientific Computing Center
(NERSC), which is supported by the US DOE Office of Science 
under Contract No. DE-AC02-05CH11231. 

\newpage
\section{Appendix}

We present tables of results that correspond to the $N_{\rm max}=400$ case
unless otherwise specified, where all results have converged to six or more digits 
of precision as a function of $N_{max}$.
All results involving the LENPIC interactions in the Hamiltonian correspond to adopting the 
LENPIC regulator $R=1.0$ fm.
Results are presented at each value of the HO basis parameter $\hbar\Omega$ and for
each observable.
Differences between the ``exact'' results in the sixth significant 
figure at different values of $\hbar\Omega$ in Tables 1 and 2 arise from the transformation 
of the HO basis representation from an initial representation at $\hbar\Omega = 28$ MeV 
where 5 significant figures in the gs energy was adopted as the criteria for accuracy in the 
HO basis transformation.
On the other hand, the results shown here to six significant figures are the results reproduced by
the OLS transformation to at least this level of accuracy.

\begin{table}[ht]
\begin{centering}
\begin{tabular}{|c|c|c|c|}
\cline{2-4} 
\multicolumn{1}{c|}{} & \multicolumn{3}{c|}{$\hbar\Omega\,{\rm (MeV)}$}\tabularnewline
\hline 
Potential & 5 & 10 & 20\tabularnewline
\hline 
\hline 
$\mbox{LENPIC--NLO}$ & -2.20607 & -2.20609 & -2.20609\tabularnewline
\hline 
$\mbox{LENPIC--N\ensuremath{^{2}}LO}$ & -2.23508 & -2.23516 & -2.23516\tabularnewline
\hline 
$\mbox{LENPIC--N\ensuremath{^{3}}LO}$ & -2.22324 & -2.22326 & -2.22326\tabularnewline
\hline 
$\mbox{Idaho--N\ensuremath{^{3}}LO}$ & -2.22458 & -2.22459 & -2.22458\tabularnewline
\hline 
\end{tabular}
\par\end{centering}

\caption{Groundstate eigenvalues (in MeV) for the specified potentials used 
as the ``exact'' values in Fig. \ref{fig:Heff}. } 
\end{table}

\begin{table}[ht]
\begin{centering}
\begin{tabular}{|c|c|c|c|}
\cline{2-4} 
\multicolumn{1}{c|}{} & \multicolumn{3}{c|}{$\mbox{\ensuremath{\hbar\Omega}\,(MeV)}$}\tabularnewline
\hline 
gs Observable & 5 & 10 & 20\tabularnewline
\hline 
\hline 
H (MeV) & -2.23508 & -2.23516 & -2.23516\tabularnewline
\hline 
$\mbox{\ensuremath{r_{\rm rms}} (fm)}$  & 1.96440 & 1.96436 & 1.96436\tabularnewline
\hline 
$\mbox{Q (e\,fm$^2$)}$ & 0.269862 & 0.269874 & 0.269873\tabularnewline
\hline 
$\mbox{\ensuremath{\mu}\ (\ensuremath{\mu_{N}})}$ & 0.856323 & 0.856323 & 0.856323\tabularnewline
\hline 
\end{tabular}
\par\end{centering}

\caption{Groundstate eigenvalues and selected observables used as the ``exact'' values
in Fig. \ref{fig:Deuteron_obervables}. The results were obtained with the 
LENPIC--N$^2$LO interaction with regulator $R=1.0$ fm. 
No confining interaction was included. }
\end{table}

\begin{table}[ht]
\begin{centering}
\begin{tabular}{|c|c|c|c|}
\cline{2-4} 
\multicolumn{1}{c|}{} & \multicolumn{3}{c|}{$\mbox{\ensuremath{\hbar\Omega}\,(MeV)}$}\tabularnewline
\hline 
gs Observable & 5 & 10 & 20\tabularnewline
\hline 
\hline 
H (MeV) & -0.703487 & 2.35148 & 10.7332\tabularnewline
\hline 
$\mbox{\ensuremath{r_{\rm rms}} (fm)}$  & 1.44078 & 1.20869 & 0.992923\tabularnewline
\hline 
$\mbox{Q (e\,fm$^2$)}$ & 0.204165 & 0.164676 & 0.122359\tabularnewline
\hline 
$\mbox{\ensuremath{\mu}\ (\ensuremath{\mu_{N}})}$ & 0.852184 & 0.849597 & 0.84814\tabularnewline
\hline 
\end{tabular}
\par\end{centering}

\caption{Groundstate eigenvalues and selected observables used as the ``exact'' values
in Fig. \ref{fig:staticMoments}. The results were obtained with the 
LENPIC--N$^2$LO interaction with regulator $R=1.0$ fm. 
The strength of the confining HO potential is the 
same as the basis parameter $\hbar\Omega$ that labels each column of results.
}
\end{table}

\begin{table}[ht]
\begin{centering}
\begin{tabular}{|c|c||c|c|c|}
\cline{3-5} 
\multicolumn{2}{c|}{} & \multicolumn{3}{c|}{$\mbox{\ensuremath{\hbar\Omega}\,(MeV)}$}\tabularnewline
\hline 
Decay & LENPIC & 5 & 10 & 20\tabularnewline
\hline 
\hline 
\multirow{2}{*}{GT} & $\mbox{NLO}$ & -1.40355 & -1.42839 & -1.43974\tabularnewline
\cline{2-5} 
 & $\mbox{N\ensuremath{^{2}}LO}$ & -1.40338 & -1.42902 & -1.44106\tabularnewline
\hline 
\multirow{2}{*}{$\mbox{0\ensuremath{\nu}2\ensuremath{\beta}}$} & $\mbox{NLO}$ & 1.59067 & 0.505287 & 0.827882\tabularnewline
\cline{2-5} 
 & $\mbox{N\ensuremath{^{2}}LO}$ & 1.48274 & 0.476412 & 0.792684\tabularnewline
\hline 
\end{tabular}
\par\end{centering}

\caption{Groundstate transition matrix elements used as the ``exact'' values in
Figs. \ref{fig:transitions} and \ref{fig:quenching}. 
The strength of the confining HO potential is the 
same as the basis parameter $\hbar\Omega$ that labels each column of results.
The GT transition matrix element values correspond to the $N_{\rm max}=400$ case, where
they have converged to six or more significant digits. 
The $\mbox{0\ensuremath{\nu}2\ensuremath{\beta}}$
transition matrix element values correspond to the $N_{\rm max}=200$ case, where
they have converged to four or more significant digits.}
\end{table}


\newpage
\begin{thebibliography}{99}

\bibitem{Okubo:1954zz} 
  S.~Okubo,
  Prog.\ Theor.\ Phys.\  {\bf 12}, 603 (1954).
  
  
\bibitem{Suzuki:1980yp} 
  K.~Suzuki and S.~Y.~Lee,
  Prog.\ Theor.\ Phys.\  {\bf 64}, 2091 (1980).
  
\bibitem{Suzuki:1982}   
  K.~Suzuki,  Prog.\ Theor.\ Phys.\  {\bf 68}, 246 (1982).

\bibitem{Stetcu:2004wh} 
  I.~Stetcu, B.~R.~Barrett, P.~Navratil and J.~P.~Vary,
  Phys.\ Rev.\ C {\bf 71}, 044325 (2005).

\bibitem{Stetcu:2006zn} 
  I.~Stetcu, B.~R.~Barrett, P.~Navratil and J.~P.~Vary,
  Phys.\ Rev.\ C {\bf 73}, 037307 (2006).
  
\bibitem{Lisetskiy:2009sh} 
  A.~F.~Lisetskiy, M.~K.~G.~Kruse, B.~R.~Barrett, P.~Navratil, I.~Stetcu and J.~P.~Vary,
  Phys.\ Rev.\ C {\bf 80}, 024315 (2009).
    
\bibitem{Barrett:2013nh} 
  B.~R.~Barrett, P.~Navratil and J.~P.~Vary,
  Prog.\ Part.\ Nucl.\ Phys.\  {\bf 69}, 131 (2013).

\bibitem{Dikmen:2015tla} 
  E.~Dikmen, A.~F.~Lisetski, B.~R.~Barrett, P.~Maris, A.~M.~Shirokov and J.~P.~Vary,
  Phys.\ Rev.\ C {\bf 91}, no. 6, 064301 (2015).

\bibitem{Anderson:2010aq} 
  E.~R.~Anderson, S.~K.~Bogner, R.~J.~Furnstahl and R.~J.~Perry,
  Phys.\ Rev.\ C {\bf 82}, 054001 (2010).

\bibitem{Epelbaum:2014sza} 
  E.~Epelbaum, H.~Krebs and U.~G.~Meißner,
  Phys.\ Rev.\ Lett.\  {\bf 115}, no. 12, 122301 (2015).
  
\bibitem{Epelbaum:2014efa} 
  E.~Epelbaum, H.~Krebs and U.~G.~Meißner,
  Eur.\ Phys.\ J.\ A {\bf 51}, no. 5, 53 (2015).


\bibitem{Maris:2016wrd} 
  P.~Maris {\it et al.},
  EPJ Web Conf.\  {\bf 113}, 04015 (2016).

\bibitem{Binder:2015mbz} 
  S.~Binder {\it et al.} [LENPIC Collaboration],
  Phys.\ Rev.\ C {\bf 93}, no. 4, 044002 (2016).

\bibitem{Binder:2018pgl} 
  S.~Binder {\it et al.} [LENPIC Collaboration],
  Phys.\ Rev.\ C {\bf 98}, no. 1, 014002 (2018).

\bibitem{Tilley:2002vg} 
D.~R.~Tilley, C.~M.~Cheves, J.~L.~Godwin, G.~M.~Hale, H.~M.~Hofmann, J.~H.~Kelley, C.~G.~Sheu and H.~R.~Weller,
Nucl.\ Phys.\  A {\bf 708}, 3 (2002).

\bibitem{Shin:2016poa} 
  I.~J.~Shin, Y.~Kim, P.~Maris, J.~P.~Vary, C.~Forssen, J.~Rotureau and N.~Michel,
  J.\ Phys.\ G {\bf 44}, no. 7, 075103 (2017).

\bibitem{Viazminsky:2001_JPV}
C.P. Viazminsky and J.P. Vary, 
{\it J. Math. Phys.} \textbf{42}, 2055 (2001).

\bibitem{Navratil:2000ww} 
  P.~Navratil, J.~P.~Vary and B.~R.~Barrett,
  Phys.\ Rev.\ Lett.\  {\bf 84}, 5728 (2000).


\bibitem{Entem:2003ft} 
  D.~R.~Entem and R.~Machleidt,
  Phys.\ Rev.\ C {\bf 68}, 041001 (2003).

\bibitem{BohrMottlesonVol1}
A.~Bohr and B.~Mottelson, Nuclear Structure, Vol. 1 (World
Scientific, Singapore, 1998), p. 407.

\bibitem{MartinezPinedo:1996vz} 
  G.~Martinez-Pinedo, A.~Poves, E.~Caurier and A.~P.~Zuker,
  Phys.\ Rev.\ C {\bf 53}, no. 6, R2602 (1996).

\bibitem{Prezeau:2003xn} 
  G.~Prezeau, M.~Ramsey-Musolf and P.~Vogel,
  Phys.\ Rev.\ D {\bf 68}, 034016 (2003).

\end{thebibliography}
\end{document}